\documentclass[sn-mathphys-num,iicol]{sn-jnl}

\usepackage{graphicx}%
\usepackage{multirow}%
\usepackage{amsmath,amssymb,amsfonts}%
\usepackage{amsthm}%
\usepackage{mathrsfs}%
\usepackage[title]{appendix}%
\usepackage{xcolor}%
\usepackage{textcomp}%
\usepackage{manyfoot}%
\usepackage{booktabs}%
\usepackage{algorithm}%
\usepackage{algorithmicx}%
\usepackage{algpseudocode}%
\usepackage{listings}%
\usepackage{chemformula}%
\usepackage[T1]{fontenc}%
\usepackage[utf8]{inputenc}%
\usepackage{xr}%
\usepackage{ulem}%
\usepackage{comment}%
\usepackage{doi}%
\usepackage{pdfpages}%

\begin{document}

\title{Water is a superacid at extreme thermodynamic conditions}

\author[1]{\fnm{Thomas} \sur{Thévenet}}
\author[1]{\fnm{Axel} \sur{Dian}}
\author[1]{\fnm{Alexis} \sur{ Markovits}}
\author[3]{\fnm{Sandro} \sur{Scandolo}}
\author*[2]{\fnm{Arthur} \sur{France-Lanord}}\email{arthur.france-lanord@cnrs.fr}
\author*[1]{\fnm{Flavio} \sur{Siro Brigiano}}\email{flavio.siro$\_$brigiano@sorbonne-universite.fr}

\affil[1]{\orgdiv{Sorbonne Université, Laboratoire de Chimie Theorique, CNRS UMR 7616}, \orgaddress{\street{4 place Jussieu }, \city{Paris}, \postcode{75005}, \state{France}}}
\affil[2]{\orgdiv{Muséum National d’Histoire Naturelle, UMR CNRS 7590, Institut de Minéralogie, de Physique des Matériaux et de Cosmochimie, IMPMC, Sorbonne Université, F-75005 Paris, France}}

\affil[3]{\orgdiv{The Abdus Salam International Centre for Theoretical Physics}, \orgaddress{\street{Str. Costiera, 11}, \city{Trieste}, \postcode{34151}, \state{Italy}}}

\maketitle

\begin{abstract}
TThe chemical behavior of water under extreme pressures and temperatures lies at the heart of processes shaping planetary interiors~\cite{ross1981sky,kadobayashi2021diamond,he2022diamond,lee2011mixtures,militzer2024phase,de2023double}, influences the deep carbon cycle~\cite{rozsa2018ab,pan2020first,fowler2024mineral,li2024synthesis}, and underpins innovative high-temperature, high-pressure synthesis of materials~\cite{wu2009catalytic,lindsey2022chemistry}. Recent experiments reveal that hydrocarbons immersed in ionized water under extreme conditions transform into heavy hydrocarbons and nanodiamonds~\cite{kadobayashi2021diamond,he2022diamond,lee2011mixtures}. However, the chemistry of water at extreme conditions and its role in hydrocarbon condensation remains poorly understood. Here, using \textit{ab initio} molecular dynamics simulations with enhanced sampling techniques and machine-learning interatomic potentials, we show that increasing pressure at high temperature induces water ionization, creating a superacid-like environment~\cite{olah1969super,olah1979superacids} that favors the protonation of hydrocarbons into transient pentacoordinated carbonium ions like \ch{CH5+}. These elusive intermediates release molecular hydrogen and yield highly reactive %electrophilic
carbocations, driving hydrocarbon chain growth toward nanodiamonds. We demonstrate how the combination of water ionization and pressure-induced methane polarization leads to superacid-driven hydrocarbon chemistry, famously known at far milder conditions~\cite{olah1969super,olah1979superacids}. Our findings reveal, for the first time, a superacid aqueous regime and establish the existence of superacid chemistry under extreme conditions. Moreover, they provide a unifying reaction network that explains chemical transformations in environments such as planetary interiors and high pressure, high temperature experiments. 
\end{abstract}

 \section*{Introduction}
 
The chemistry of water under extreme conditions is of critical importance across several fields, ranging from planetary science~\cite{ross1981sky,kadobayashi2021diamond,he2022diamond,lee2011mixtures,militzer2024phase,de2023double}, terrestrial geochemistry~\cite{rozsa2018ab,fowler2024mineral,li2024synthesis} to the origin of life~\cite{li2024synthesis,goldman2010synthesis}, as well as for the high-pressure and high-temperature (HPHT) synthesis of new materials~\cite{wu2009catalytic,lindsey2022chemistry,kapil2022first}. Within the deep interiors of planets, such as Earth’s mantle or the fluid layers of icy giants and sub-Neptune exoplanets, water and carbon based compounds are subjected to high pressures and temperatures, driving unique chemical transformations.
As pressure and temperature rise, water undergoes increased ionization~\cite{goncharov2005dynamic,cheng2021phase,schwegler2001dissociation,cavazzoni1999superionic}, creating a highly reactive environment that profoundly impacts the chemical stability and reactivity of dissolved compounds~\cite{wu2009catalytic,pan2013dielectric,huang2023formation,li2024synthesis}. For instance, under conditions typical of Earth's upper mantle (10 GPa, 1000--1745 K), the pronounced ionization of water stabilizes \ch{HCO3-} species over \ch{CO2}, effectively excluding the presence of \ch{CO2} in water-rich geofluids in the Earth~\cite{pan2020first}. 

One of the most enigmatic transformations under extreme conditions in water is the polycondensation of simple hydrocarbons into heavier hydrocarbons and nanodiamonds. The HPHT synthesis of heavy hydrocarbon mixtures and diamonds carries significant implications for planetary science~\cite{ross1981sky,benedetti1999dissociation,ancilotto1997dissociation,kadobayashi2021diamond,he2022diamond,lee2011mixtures,kraus2017formation,cheng2023thermodynamics,ghiringhelli2004high,frost2024diamond,roy2024mixture}, geochemistry~\cite{frezzotti2019diamond,huang2023formation,fowler2024mineral}, detonation experiments~\cite{wu2009catalytic} and petrochemistry~\cite{uguna2016retardation}.
In planetary science, the “diamonds in the sky” hypothesis~\cite{ross1981sky} posits that, within the interiors of icy giants, methane and water, the primary constituents of their fluid mantles, transform into diamonds and ionized water, potentially explaining the anomalous magnetic fields and luminosity of these planets. Recent experimental evidence from Laser-Heated Diamond Anvil Cell (LHDAC)~\cite{kadobayashi2021diamond} and shock compression experiments~\cite{he2022diamond} support this hypothesis by probing the transition of C/H/O mixtures into heavy hydrocarbons, nanodiamonds, and ionized water. Despite these observations, the mechanisms governing hydrocarbon dissociation and condensation in ionized water remain largely unexplored due to the challenges of characterizing these reactions \textit{in situ}. 

Atomistic simulations have emerged as a powerful approach for
studying the chemical and physical properties of water and hydrocarbon mixtures under extreme conditions~\cite{lee2011mixtures,de2023double,cheng2023thermodynamics,militzer2024phase,ancilotto1997dissociation,conway2021rules,cassone2021molecular}. Lee \textit{et al.}~\cite{lee2011mixtures} and Militzer~\cite{militzer2024phase} conducted \textit{ab initio} simulations respectively on \ch{H2O}/\ch{CH4} and \ch{H2O}/\ch{CH4}/\ch{NH3} mixtures under conditions relevant to icy giant interiors: they both report the formation of C--O and C--C bonds, promoted by increases in temperature or pressure. 
Despite these observations, the fundamental chemistry governing hydrocarbon transformations in water under extreme conditions, as well as the interplay between pressure, degree of water ionization and hydrocarbon chain elongation, remain \textit{terra incognita}.
In this study, we investigate the chemical behavior of water/methane mixtures under extreme conditions by simulating their reactivity in pressure-temperature regimes where heavy hydrocarbon and diamond formation have been experimentally observed~\cite{kadobayashi2021diamond,he2022diamond}. First, using a combination of DFT-MD and machine-learning interatomic potentials (MLIPs), we explore methane condensation pathways in water. Our simulations reveal that water, in such extreme conditions, is capable of protonating methane and behaves therefore as a superacid, leading to the formation of non-classical \ch{CH5+} ions. We show that such a superacidic regime of water in extreme conditions catalyzes methane polycondensation via transient pentacoordinated carbonium ions (e.g., \ch{CH5+}, \ch{CH3CH4+}). The chemistry we observe taking place closely mirrors the superacid-catalyzed condensation of hydrocarbons at moderate temperatures (60--150$^\circ$C) and ambient pressure, as described in the Nobel Prize-winning studies of Olah \textit{et al.}~\cite{olah1969super,olah1979superacids}. Then, using enhanced sampling techniques, we investigate chain elongation, hydrocarbon branching, and diamond-like structure formation, which eventually allows us to sketch a comprehensive framework for understanding how high pressure and water ionization drive such transformations. Remarkably, we find that only three elementary superacid mechanisms underpin the whole reaction network.

\section*{Superacid behavior of water}\label{sezioneone}

\begin{figure*}[h!]
\includegraphics[width=1.0\textwidth]{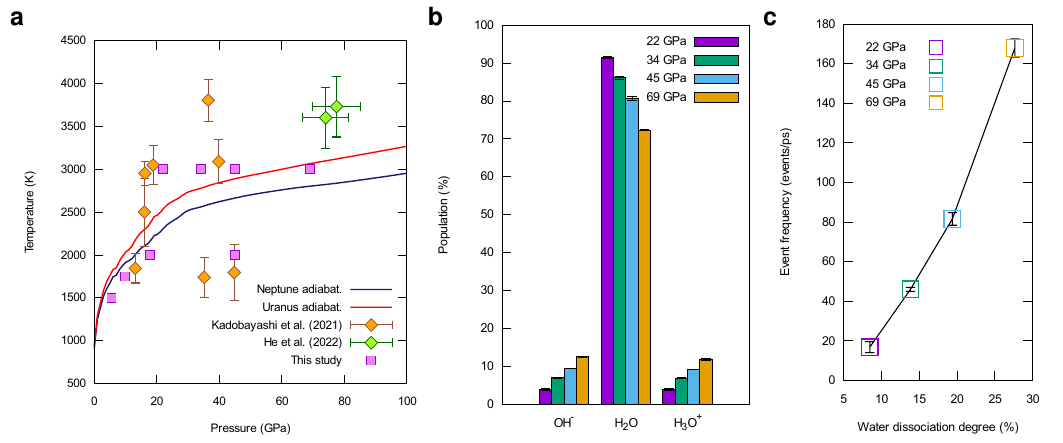}
\caption{\textbf{Panel a}: Pressure and temperature conditions explored in this study (purple squares), alongside previous LH-DAC~\cite{kadobayashi2021diamond} (orange rhombuses) and shock wave experiments~\cite{he2022diamond} (green rhombuses) on C/H/O mixtures. The solid blue and red lines depict the predicted isentropes of Uranus and Neptune from ref.\cite{scheibe2019thermal}; \textbf{Panel b}: Population of \ch{H2O},\ch{H3O+} and \ch{OH-} species in the 22-69 GPa range at 3000K. \textbf{Panel c}: \ch{CH5+} formation frequency as a function of water dissociation degree. \textbf{For Panel b,c}: error bars correspond to a 95\% confidence interval. \label{fig:def1}}
\end{figure*}

We simulated \ch{H2O/CH4} mixtures using simulation boxes composed of 76 water molecules and 52 methane (488 atoms), a stoichiometry approximately close to the protosolar ratio of heavy nuclei~\cite{lodders2010solar}. Eight P--T conditions were investigated (purple squares in Fig.~\ref{fig:def1}a) that closely approach the adiabats of Uranus and Neptune. For all conditions explored in this work, the mixture behaves as a dense liquid characterized by a diffusive behavior of heavy atoms (see Supplemental Table 1). Observations from our DFT-MD trajectories revealed a specific pressure--temperature region (22--69 GPa at 3000 K) where water becomes significantly ionized, dissociating into hydronium (\ch{H3O+}) and hydroxide (\ch{OH-}) ions. In this region the water dissociation monotonously increases as a function of pressure, with values ranging from 8.5 $\%$ at 22 GPa to 27.7 $\%$ at 69 GPa. This can be easily appreciated in Fig.~\ref{fig:def1}b where the water species populations for all of the four Pressure conditions at 3000 K are reported. The same effect is characterized for pure water, as reported in Supplementary Fig. 1. 

Under these conditions, frequent proton hopping events among water molecules in the mixture occur along the dynamics, following the Grotthuss mechanism. Notably, methane molecules also participate in these proton hopping chains, becoming protonated by either water molecules or hydronium ions, which leads to the formation of transient methanium cations (\ch{CH5+}).

This corresponds to a superacid behavior of water in extreme conditions, able to protonate extremely weak bases such as methane~\cite{olah1969super,olah1979superacids}. We evaluated the superacidity of the mixture across the 22--69 GPa pressure range by computing the frequency of \ch{CH5+} formation at 3000 K as a function of the water dissociation degree (see Fig.~\ref{fig:def1}c). The rate is defined as the number of \ch{CH5+} formation events per picosecond. The criteria establishing these events are detailed in the Methods section.
The analysis presented in Fig.~\ref{fig:def1}c highlights a direct correlation between the \ch{CH5+} rate of formation and the degree of water dissociation. As water dissociates under the effect of growing pressure, the \ch{CH5+} formation frequency increases almost linearly.

\begin{figure*}
\includegraphics[width=1\textwidth]{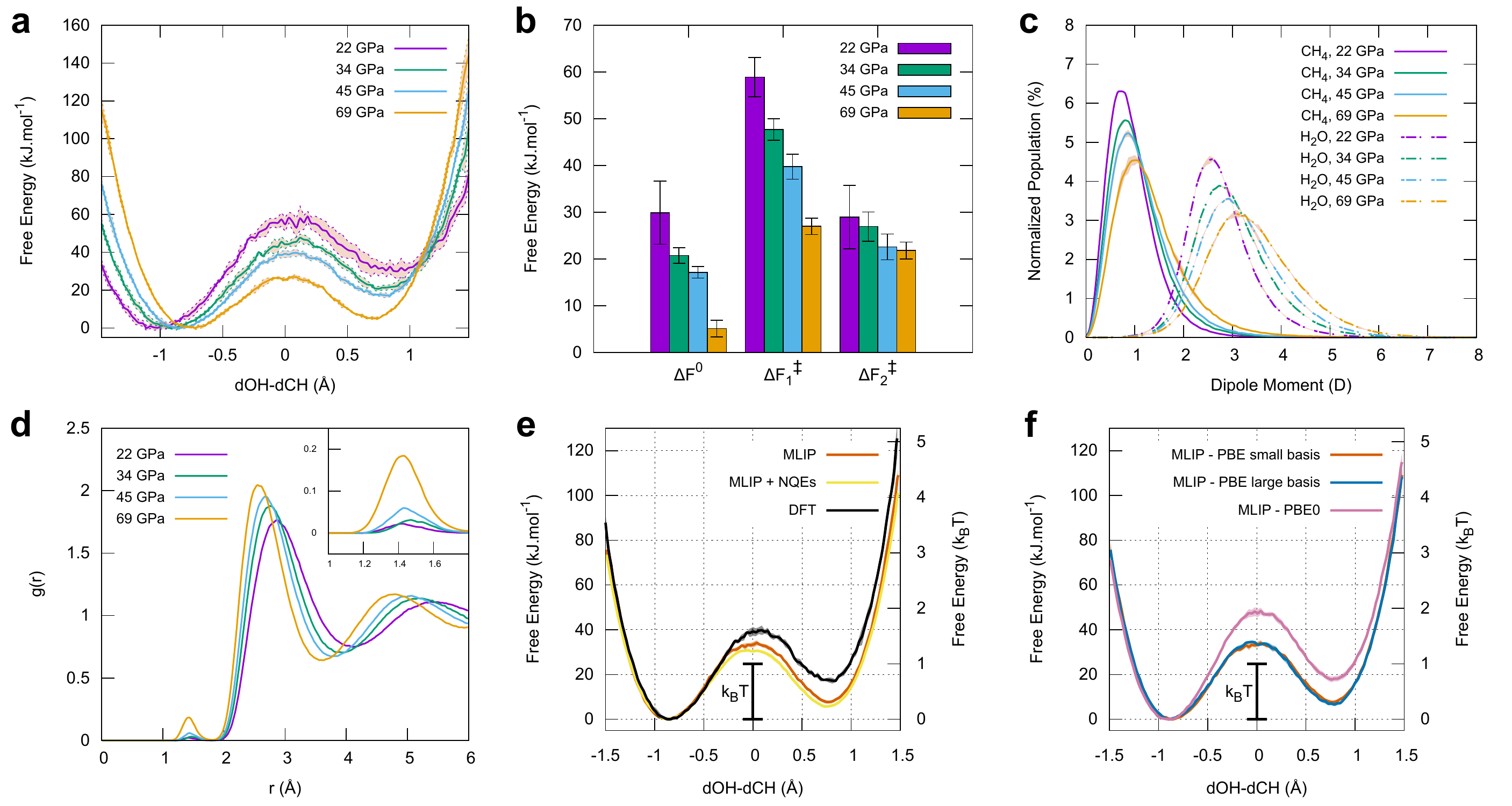}
\caption{\textbf{Panel a:} Free energies of \ch{CH5+} formation as a function of the reaction coordinate, defined as $d(\text{O--H}) - d(\text{C--H})$, the difference between the distance from the hydrogen to the nearest oxygen and the distance from the same hydrogen to the nearest carbon atom.
 \textbf{Panel b:} Histogram plot showing the free energy barriers ($\Delta F_1^{\ddagger}$ for the forward reaction, $\Delta F_2^{\ddagger}$ for the backward reaction) and the reaction free energy ($\Delta F$). \textbf{Panel c:} Population of the molecular dipole moments of \ch{CH4} and \ch{H2O} in the 22-64 GPa range at 3000K. \textbf{Panel d:} Combined C--O and C--C radial distribution functions in the 22-64 GPa range at 3000K. \textbf{Panel e:} Comparison between free energy profiles obtained through DFT-MD (black line), and with a surrogate MLIP model through classical molecular dynamics (red line) or with the inclusion of nuclear quantum effects (yellow line). \textbf{Panel f:} Free energy profiles comparing different MLIPs trained on three datasets: DFT-MD settings ("PBE small basis", red line), larger basis set ("PBE large basis", blue line), and a hybrid functional ("PBE0", pink line). \textbf{For Panels a,b,c,e,f}: All values are accompanied by an error bar corresponding to a 95$\%$ confidence interval.\label{fig:def2}}
\end{figure*}

To rationalize the thermodynamics and kinetics of this phenomenon, we present in Fig.~\ref{fig:def2}a-b the converged free energy profiles from unbiased DFT-MD, associated with the protonation of methane by a hydronium ion (Eq.\ref{eq_ch51}) at 3000 K across the pressure range of interest. 

\begin{equation}
\label{eq_ch51}
\ch{H_3O^+} + \ch{CH_4} \rightleftharpoons \ch{H_2O} + \ch{CH_5^+}
\end{equation}

From 22 to 69 GPa, the protonation of \ch{CH4} becomes kinetically favored, as evidenced by an approximate 50\% reduction in the free energy barriers ($\Delta F_1^{\ddagger}$). By contrast, the barrier for the \ch{CH5+} decomposition ($\Delta F_2^{\ddagger}$) is only slightly affected by a change in pressure, which is consistent with the invariance of the \ch{CH5+} lifetimes with respect to pressure (see Supplementary Fig. 2).
Interestingly, the free energy profiles also show a reduction of $\Delta F$ ($F$\(_{\text{\ch{CH4 + H3O+}}}\) - $F$\(_{\text{\ch{CH5+ + H2O}}}\)) by more than threefold: the reaction is therefore thermodynamically favored with pressure increase. Note that the molecular recognition analysis implemented in this study to compute the free energies of protonation (Fig.~\ref{fig:def2}a--b) and for population analysis (Fig.~\ref{fig:def1}b) only weakly depends on the chosen O--H and C--H cutoffs (see the Methods section for more details). As reported in Fig.~\ref{fig:def2}e-f and discussed in greater length in Supplementary Section 9, using surrogate models in the form of machine-learned interatomic potentials~\cite{musaelian2023}, we have verified that neither the generalized gradient approximation and basis set size nor nuclear quantum effects significantly impact the free energy profiles we report. All reported errors lie within a fraction of $k_BT$ with respect to $\Delta F$ and $\Delta F_1^{\ddagger}$. The free energy profiles associated to the \ch{CH5+} formation from water and methane (see Eq.\ref{eq_ch52}) are reported in Supplementary Fig. 3. 

\begin{equation}
\label{eq_ch52}
\ch{H_2O} + \ch{CH_4} \rightleftharpoons \ch{OH^-} + \ch{CH_5^+}
\end{equation}

In spite of more than twice higher free energy barriers with respect to the reaction from methane and hydronium (see Eq.\ref{eq_ch51}), the trends with pressure are qualitatively the same. However, the large difference in $\Delta F$ between the two channels of \ch{CH5+} formation makes the pathway through Eq.\ref{eq_ch51} the most favorable one, irrespective of the pressure.

As pressure increases, the observed reduction in free energy barriers for both reaction channels aligns with the corresponding rise in the dipole moment of methane. This can be appreciated in Fig.~\ref{fig:def2}c, where the instantaneous dipole moment of methane, calculated using Wannier centers~\cite{marzari1997maximally}, is presented at 3000 K within the pressure range of 22-69 GPa. The pressure-induced deviation from ideal tetrahedral coordination results in methane molecules exhibiting an instantaneous dipole moment, with the maximum absolute value progressively increasing from 0.73 D at 22 GPa to 1.00 D at 69 GPa. Additionally, with rising pressure, the distributions widen, indicating a growing number of methane molecules with elevated dipole moments in the simulation. The dipole moment induced by pressure enhances the basic character of \ch{CH4} molecules, thereby favoring their protonation. The influence of water polarization and ionization on methane’s acquired dipole moment is detailed in Supplementary Section 4. 
 
We conclude that pressure plays a dual role in promoting the methane protonation reaction: it both lowers the free energy barrier and the reaction free energy toward \ch{CH5+} formation, and increases the concentrations of \ch{H3O+}, the reactant of the most favorable methane protonation pathway (Eq. \ref{eq_ch51}).

In Supplementary Section 5, we report the observation of the same superacid chemistry at dilute solution conditions (1 methane for 127 water molecules) and lower temperature (2000 K), including the formation of non-classical \ch{CH5+} species. This establishes the existence of water superacidity in a large P,T range and over many \ch{CH4} concentrations. 

\section*{Extreme superacid-catalyzed condensation of hydrocarbons}\label{sezionetwo}

\begin{figure*}[h!]
\includegraphics[width=1.0\textwidth]{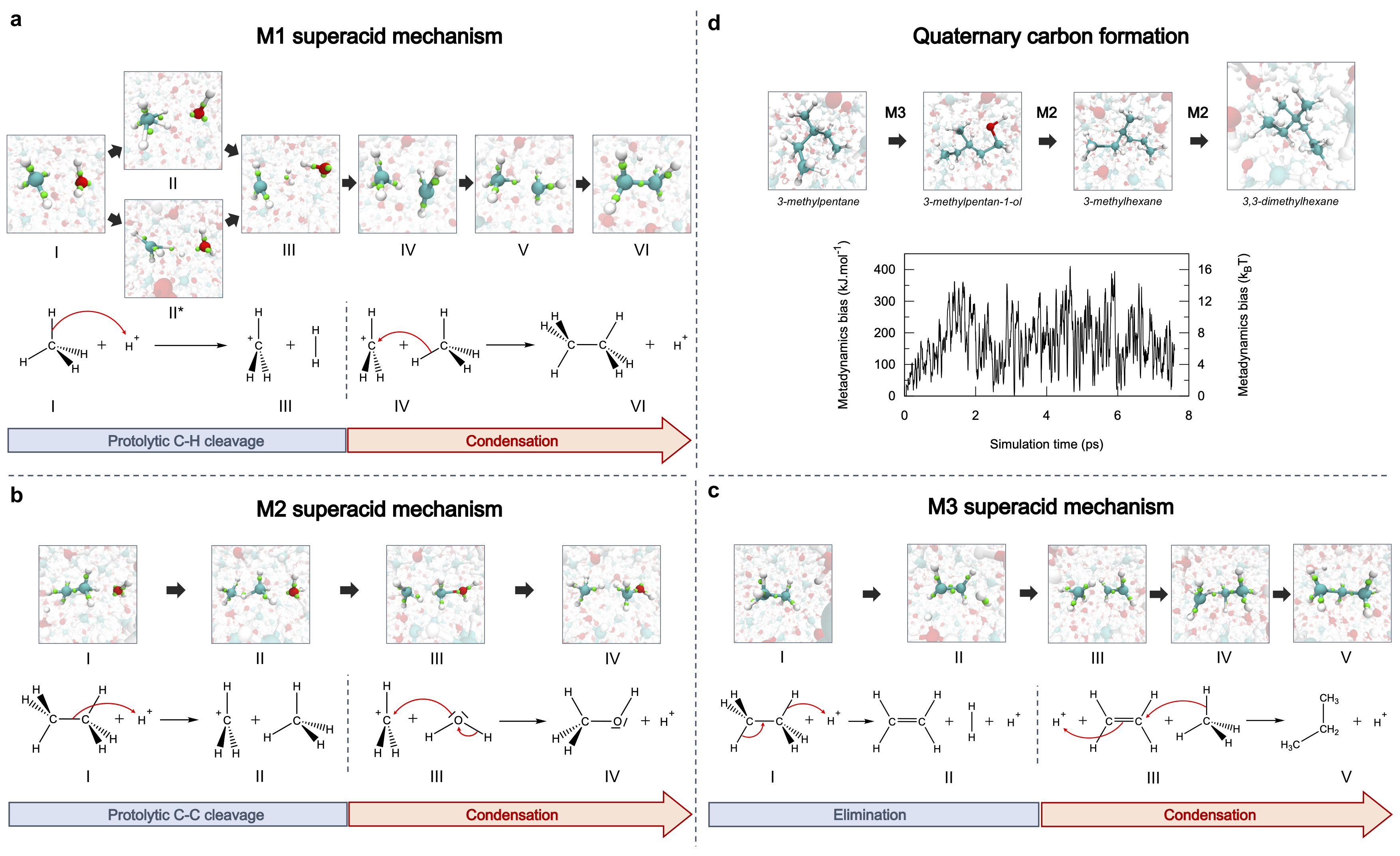}
\caption{Scheme of the reaction mechanisms, along with the Wannier centers associated with the reactive structures for the superacid M1 (\textbf{Panel a}) and M2 (\textbf{Panel b}), as well as the M3 superacid catalyzed reaction mechanism (\textbf{Panel c}). \textbf{Panel d}: scheme of the reaction mechanism (top) and associated metadynamics bias (bottoms) for diamond-like structure formation at 3000K and 50 GPa.}
\label{fig:def3}
\end{figure*}

We have then focused on the formation of C--O and C--C covalent bonds that occurs spontaneously along the DFT-MD simulations at 3000 K within the pressure range of 22 to 69 GPa. This is clearly illustrated by the combined C--O and C--C radial distribution functions shown in Fig.~\ref{fig:def2}d. The emergence of a peak at a short distance (approximately 1.4 \AA), which increases in intensity with rising pressure, indicates the formation of C--C and C--O bonds throughout the dynamics. Methanol and ethane are formed across the entire pressure range, while the formation of longer hydrocarbons (propane and butane) is observed only at 45 and 69 GPa, respectively. We find the emergence of pentacoordinated carbonium ions (e.g. \ch{CH5+}, \ch{R-CH4+}), to be one of the molecular keys to rationalize the hydrocarbon chain elongation in such conditions. For instance, the DFT-MD simulations reveal that once formed, the \ch{CH5+} ion either undergoes reversible deprotonation, accounting for hydrogen exchange, or loses molecular hydrogen to form the highly reactive carbonium ion \ch{CH3+} (structure III in Fig.~ \ref{fig:def3}a). The mechanism, along with the Wannier centers associated with the reactive structures, is shown in Fig.~\ref{fig:def3}a. The reaction is initiated by the protonation of methane, resulting in a metastable \ch{CH5+} structure characterized by a two-electron, three-center C--H${_2}$ bond (structure II). The reaction proceeds with the cleavage of the C--H${_2}$ bond, release of \ch{H2} into the solution and formation of the highly reactive \ch{CH3+} ion. Finally, ethane or methanol molecules are formed through an electrophilic attack of the \ch{CH3+} ion on a neighboring methane or water molecule, along with the concurrent release of a proton. We will henceforth refer to such reaction as the M1 superacid mechanism. As observed in the DFT-MD simulations, the same mechanism also leads to the formation of propane (45 GPa) and butane (69 GPa). In these cases, the reaction proceeds either through the formation of \ch{CH5+} or more complex pentacoordinated carbonium ions (\ch{CH3CH4+},\ch{CH3CH2CH4+}); their dehydrogenation follows, as well as the formation of highly reactive trivalent carbenium ions (\ch{CH3CH2+},\ch{CH3CH2CH2+}). Over the whole pressure range, most of the C--C and C--O bonds are stable for a large portion of the simulations. It is important to note that the three-center two-electron bond identified for the \ch{CH5+} in our simulations is in agreement with the gas phase \ch{CH5+} structure characterized by infrared spectroscopy and \textit{ab initio} calculations in the literature~\cite{huang2006quantum,marx1999ch5+}. 

The M1 mechanism can also proceed through the formation of a slightly different \ch{CH5+} geometry (structure II$^*$), observed at 3000 K as reported in Fig.~\ref{fig:def3}a. In this case, the reactive electronic doublet is displaced close to the reactive C--H hydrogen, ready to be caught by the incoming \ch{H+}. The proton attacks the hydrogen atom of methane, leading to an unstable \ch{CH5+} geometry. 

Remarkably, the M1 superacid mechanism revealed by our simulations under extreme conditions corresponds exactly to the mechanism experimentally characterized for the oligomerization of methane in superacid solutions at moderate temperatures (60--150$^\circ$C) and ambient pressure, as reported in the landmark studies by Olah \textit{et al.}~\cite{olah1969super,olah1979superacids}. In these transformations, the condensation of methane in superacid solutions (e.g., \ch{SbF5/HF}, approximately $10^9$
times stronger than sulfuric acid) is driven by the formation of pentacoordinated carbonium ions such as \ch{CH5+} and \ch{RCH4+}, which undergo subsequent dehydrogenation. This process leads to the generation of highly reactive trivalent carbocations (e.g., \ch{CH3+}, \ch{RCH2+}) triggering condensation reactions. 

In such experimental conditions, the superacidic character of the liquid solution (\ch{SbF5/HF}), due the combined effects of a Lewis and Br{\o}nsted acid, induces the protonation of the weakly basic \ch{CH4} molecules. In stark contrast, the superacidic behavior observed in our simulations is a direct consequence of the extreme temperature and pressure conditions. The extreme conditions promote water ionization, significantly increasing the concentration of \ch{H3O+}, and enhance the basicity of methane through the induction of a dipole moment. Our study demonstrates, for the first time, that superacid chemistry can occur in aqueous solutions, thereby introducing the concept of superacidity in extreme conditions ("extreme superacidity"). 

We have also identified a second superacid reaction mechanism, denoted as M2. This mechanism underlines the transitions between alcohols and alkanes, and \textit{vice versa}, as well as between different alkanes. In this regard, the spontaneous interconversion of ethane to methanol observed along the DFT-MD trajectories is characterized by Wannier centers analysis in Fig.~\ref{fig:def3}b. In this mechanism, the protolytic cleavage occurs on the C--C bonds and results in the formation of the reactive \ch{CH3+} ion. 
Interestingly, the M2 superacid mechanism has also been proposed, along with the M1 mechanism, by Olah \textit{et al.}~\cite{olah1973electrophilic} for alkane polycondensation and isomerization in superacid solutions. This parallel further confirms the superacidic nature of hydrocarbon chemistry in such conditions.

Finally, we identified a third mechanism, denoted as M3. This elongation reaction involves the formation of double-bond intermediates through an initial elimination step, followed by a condensation reaction (see Fig.~\ref{fig:def3}c). In this process, the double-bonded species acts as an electrophilic partner to a neighboring molecule. The M3 mechanism can proceed \textit{via} either a superacid-catalyzed or a base-catalyzed pathway. For example, Fig.~\ref{fig:def3}c illustrates the formation of propane through the M3 superacid pathway. The description of the M3 base-catalyzed mechanism and lifetime analysis of the double bond intermediate are described in Supplementary Section 6. 

\section*{Hydrocarbon branching and diamond-like structure formation}\label{sezionefour}

 \begin{figure*}[h!]
 \center
\includegraphics[width=0.7\textwidth]{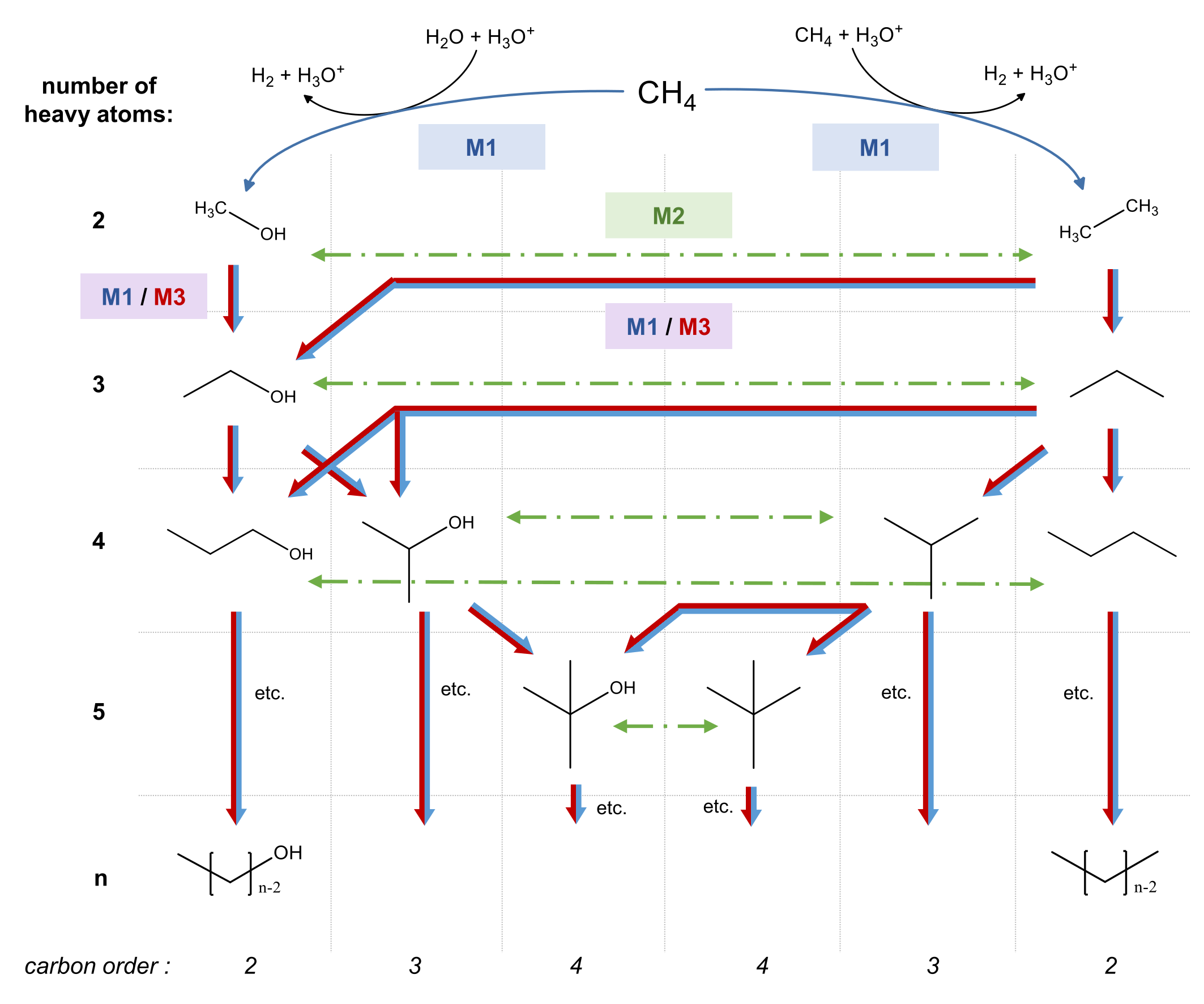}
\caption{Scheme of the reaction network for the transformation of methane into nanodiamond-like structures in high-pressure, high-temperature water. Rows: number of carbon and oxygen atoms in molecular compounds. Columns: maximum carbon substitution degree within a molecular compound (\textit{i.e.} secondary, tertiary, quaternary carbons). Blue arrows: elongation reactions via the M1 mechanism. Blue/Red arrows: elongation and branching reactions through either the M1 or M3 mechanisms. Green arrows: alcohol/alkane interconversion reactions through the M2 mechanism.}
\label{fig:def4}
\end{figure*}

In the previous sections, we provided a mechanistic rationalization of the early stages of hydrocarbon chain elongation in water. Here, we shift our focus to the reactivity in later stages, specifically hydrocarbon branching and diamond-like carbon structure formation. To achieve this, we performed metadynamics simulations, biaising along a topologically-aware collective variable knicknamed SPRINT~\cite{pietrucci2011graph}, which is a general-purpose CV designed to explore the possible arrangements of a system's bond network (details are reported in Supplementary Section 8). Using these coordinates, we first performed a series of six independent metadynamics simulations of water/methane mixtures using systems of 488 atoms at 3000 K and 45 GPa. By examining the maximum value of the metadynamics energy bias introduced during the simulations, we obtain an upper bound of the free energy barriers associated with these transformations, thereby assessing their feasibility. On the limited timescale of the metadynamics (17 ps maximum), three of the six simulations led to the formation of complex branched hydrocarbon species containing secondary and tertiary carbon centers (\textit{i.e.} carbons bonded to two and three other carbons, respectively). In the other three simulations the \ch{H2O/CH4} mixture evolves into a mixture of simple molecules including e.g. several propane, ethanol, formic acid and ethane molecules. Crucially, all the six simulations followed the mechanism identified in the previous sections (M1, M2, and M3) and revealed free energy barriers of less than 27.8 $k_BT$. These results demonstrate the favorability of hydrocarbon chain elongation and branching, extending the superacid chemistry in water under extreme conditions discussed earlier. 

Next, we investigated the formation of diamond-like carbon structures bearing quaternary carbon atoms (\textit{i.e.} carbons bonded to four other carbons). For this purpose, we performed metadynamics on a larger system composed of 722 atoms, containing \ch{H2O/CH4} and a 3-methylpentane molecule previously obtained in our smaller-scale simulations, the detailed mechanism of its formation being reported in the Supplementary Fig. 9. As illustrated in Fig.~\ref{fig:def3}d (top panel), at 3000 K and 45 Gpa, 3-methylpentane evolved into 3,3-dimethylhexane, a quaternary carbon structure, following the superacid M3 and M2 pathways. Importantly, the metadynamics bias never exceeded 16 $k_BT$ (see Fig.~\ref{fig:def3}d , bottom panel), indicating that these transformations are energetically accessible under these conditions. Reinforcing this observation, an unbiased DFT-MD simulation initially containing a tertiary hydrocarbon species (generated by previous metadynamics) revealed the spontaneous formation of a quaternary alcohol, which subsequently interconverted into a diamond-like quaternary hydrocarbon species via a superacid M2 mechanism within a few picoseconds (see Supplementary Fig. 8). The free energy barrier associated with quaternary carbon formation at 3000 K and 45 GPa is therefore probably much lower than the estimated upper bound from metadynamics. Finally, we confirmed the metastability of the 3,3-dimethylhexane (last structure in Fig.~\ref{fig:def3}c) by performing six unbiased MD simulations. We observed that the quaternary carbon structure remains stable for more than 5.5 ps in three of these six simulations at 3000 K and 45 GPa. 

By distilling the results presented above, we are now able to present, in Fig.~\ref{fig:def4}, a general picture of hydrocarbon growth toward nanodiamond formation in HPHT water as a reaction network. Initially, \ch{CH4} condenses exclusively via the superacid M1 mechanism (blue pathways), yielding ethane or methanol. As molecular complexity increases, hydrocarbon chains extend and branch through the superacid M1 and M3 pathways (blue and red arrows), progressively forming larger structures with increasing ternary and quaternary carbon atoms. Additionally, alcohols and alkanes interconvert via the superacid M2 mechanism, further enriching molecular diversity. This framework, unveiled in our study, deciphers the stepwise transformation of methane in ionized water, revealing the core chemistry driving hydrocarbon condensation and nanodiamond formation under extreme conditions. 

\section*{Conclusions}

We show, for the first time, that water behaves as a superacid at extreme thermodynamic conditions: these findings therefore open new frontiers in the study of water at high pressure and temperature. This novel concept of "extreme superacidity" may play a dominant role in domains where water is ionized, including Earth’s upper mantle, superionic ice, and the recently discovered superionized phase of nano-confined water~\cite{kapil2022first}.
To demonstrate this phenomenon we integrated high-level electronic structure theory, density functional theory (DFT), machine learning, and advanced enhanced sampling techniques.
Our finding show that ionized water promotes the formation of pentacoordinated \ch{CH5+} carbonium ions, which release molecular hydrogen and yield highly electrophilic tricoordinated carbocations, driving hydrocarbon chain growth toward nanodiamond formation. This chemistry parallels the superacidic condensation of hydrocarbons reported in the Nobel Prize-winning studies of Olah \textit{et al}.~\cite{olah1969super,olah1979superacids}. 

Building on these insights, our framework can guide future HPHT synthesis route in ionized water, with significant implications for petrochemistry, geochemistry, and HPHT materials synthesis. Finally, these findings hold relevance for planetary science, offering key insights for developing thermodynamic and kinetic models to trace the chemical evolution of dense C/H/O mixtures in the interiors of icy giants and sub-Neptune planets. Moreover, since our results demonstrate that \ch{CH4} participates in the proton hopping network with water, this constitutes the first evidence that carbon must contribute to shallow mantle conductivity, with implications on icy giants' magnetic fields.

\begin{appendices}

\makeatletter
\renewcommand{\fnum@figure}{Extended Data Fig. \thefigure}
\makeatother

\makeatletter
\renewcommand\tablename{Extended Data Table}
\makeatother

\section*{Methods}

\subsection*{DFT: molecular dynamics}\label{methods:dftmd}

All DFT-MD simulations relied on the Born-Oppenheimer approximation, and were performed in the canonical ($NVT$) ensemble with a time step of 0.5 fs, using a Nosé-Hoover chain of three thermostats with a time constant of 50 fs. Starting configurations were generated using \textsc{packmol}\cite{martinez2009}. For each P--T conditions, two independent dynamics ranging from 35 to 42 ps have been performed using simulation boxes of 488 atoms (52 \ch{CH4} and 76 \ch{H2O} molecules). All the analysis and associated error bars are performed on these trajectories, excluding the first 7 ps of equilibration. Larger simulation boxes (722 atoms) were adopted to probe the stability and reactivity of complex hydrocarbon species featuring tertiary and quaternary carbon centers. 

All DFT-MD simulations were performed using the \textsc{CP2K} package\cite{vandevondele2005quickstep}, under the generalized gradient approximation (GGA), with the PBE functional\cite{perdew1996generalized} and D3 Becke-Johnson dispersion corrections\cite{grimme2011effect}. We solved the spin-restricted Kohn-Sham equations self-consistently using the orbital transformation method, with an associated convergence criterion set to $10^{-6}$ Ha. The electronic density and wavefunction are represented using a dual basis of atomic orbitals and plane waves, as implemented in \textsc{quickstep}\cite{vandevondele2005quickstep}; we truncate the plane wave expansion using a 950 Ry cutoff, and use DZVP-MOLOPT-SR basis sets. 

Metadynamics-accelerated DFT-MD simulations have been performed in order to enhance the sampling of the \ch{H2O}/\ch{CH4} mixture phase space toward hydrocarbon branching and diamond-like structure formation. Details regarding metadynamics simulations and the definition of the SPRINT reaction coordinates are presented in Supplementary Section 8.

\subsection*{Molecular recognition method}\label{recognition}

The molecular recognition method adopted for the population analysis (Fig.~\ref{fig:def1}b) and free energy profiles (Fig.~\ref{fig:def2}a,e,f), is based on the postulate that each hydrogen atom is singly bonded to one heavy atom (carbon or oxygen). Each hydrogen atom is assigned to its first neighboring heavy atom (C or O). This procedure successfully identifies all the relevant molecular species (\ch{CH4}, \ch{CH5+}, \ch{H3O+},\ch{OH-}, \ch{CH3+}).
However, the possible presence of \ch{H2} molecules in the system, formed as a byproduct of C--C and C--O bond formation reactions, necessitates the introduction of a radial cutoff to exclude these atoms from recognition. To achieve this, we apply radial cutoffs to H--H, C--H, and O--H distances. Specifically, we identify hydrogen atoms that do not fall within the coordination shells of carbon and oxygen, where the shell radii are defined by the respective C--H and O--H cutoffs. Next, we determine whether these hydrogen atoms are linked to other hydrogen atoms that are not associated with heavy atoms, within the proximity defined by the H--H cutoff. If no H--H linkage is detected, the hydrogen atom is assigned to the nearest heavy atom.
We assessed the sensitivity of our analysis to the choice of C--H and O--H cutoff values, as illustrated in Extended Data Fig. \ref{fig:cutoff} for four systems at 3000 K in the pressure range of 22--69 GPa. The cutoff values fall within a stable plateau region, confirming the robustness of our method and its independence from the specific C--H and O--H cutoff values. To further validate this, we performed our analysis on parts of the trajectory that do not contain \ch{H2} molecules and compared the results obtained with and without the cutoff. In both cases, the results remained consistent, demonstrating the reliability of our approach.\\
As a final test, we assessed the performance of the molecular recognition operation by augmenting it with molecular charges from a Wannier center analysis. Species are therefore identified using both the coordination number of heavy atoms and the overall molecular charge. The molecular charges are determined by associating each Wannier center with its nearest neighbor (either a heavy atom or a hydrogen atom from \ch{H2} molecules). In Extended Table \ref{tbl:population} we display populations computed with and without the criterion on Wannier centers, using a 13 ps trajectory at 45 GPa and 3000 K. The excellent agreement between both approaches indicates that including Wannier centers in the molecular recognition operation is not mandatory. 

\subsection*{Calculating the \ch{CH5+} rate of formation} \label{frequency}

Here we describe the approach used to compute the frequency at which \ch{CH5+} is formed, presented in Fig.~\ref{fig:def1}c. 
Instead of focusing on formation (a proton transfer from an oxygen atom to a \ch{CH4}), we enumerate \ch{CH5+} destruction events (a proton transfer from \ch{CH5+} to an oxygen atom), which allows us to characterize in addition its lifetime distribution. 

\begin{figure*}[h!]
\center
\includegraphics[width=0.5\linewidth]{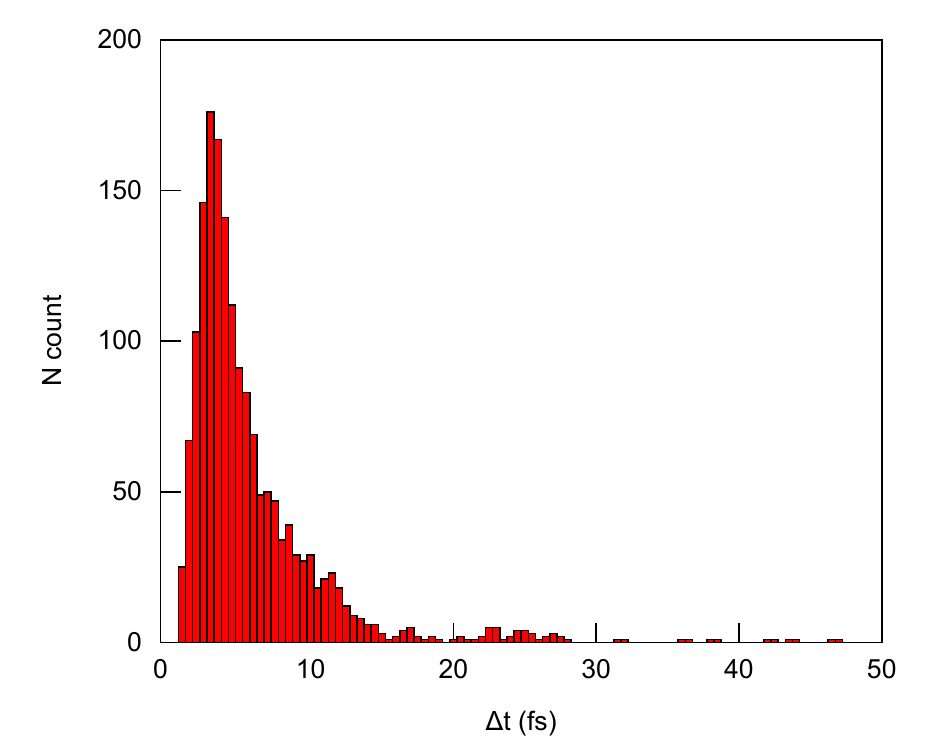}
\caption{Number of proton transfers from carbon to oxygen that are excluded by the intermittent function.}
\label{fig:tempo}
\end{figure*}

\begin{figure*}[h!]
\includegraphics[width=0.5\linewidth]{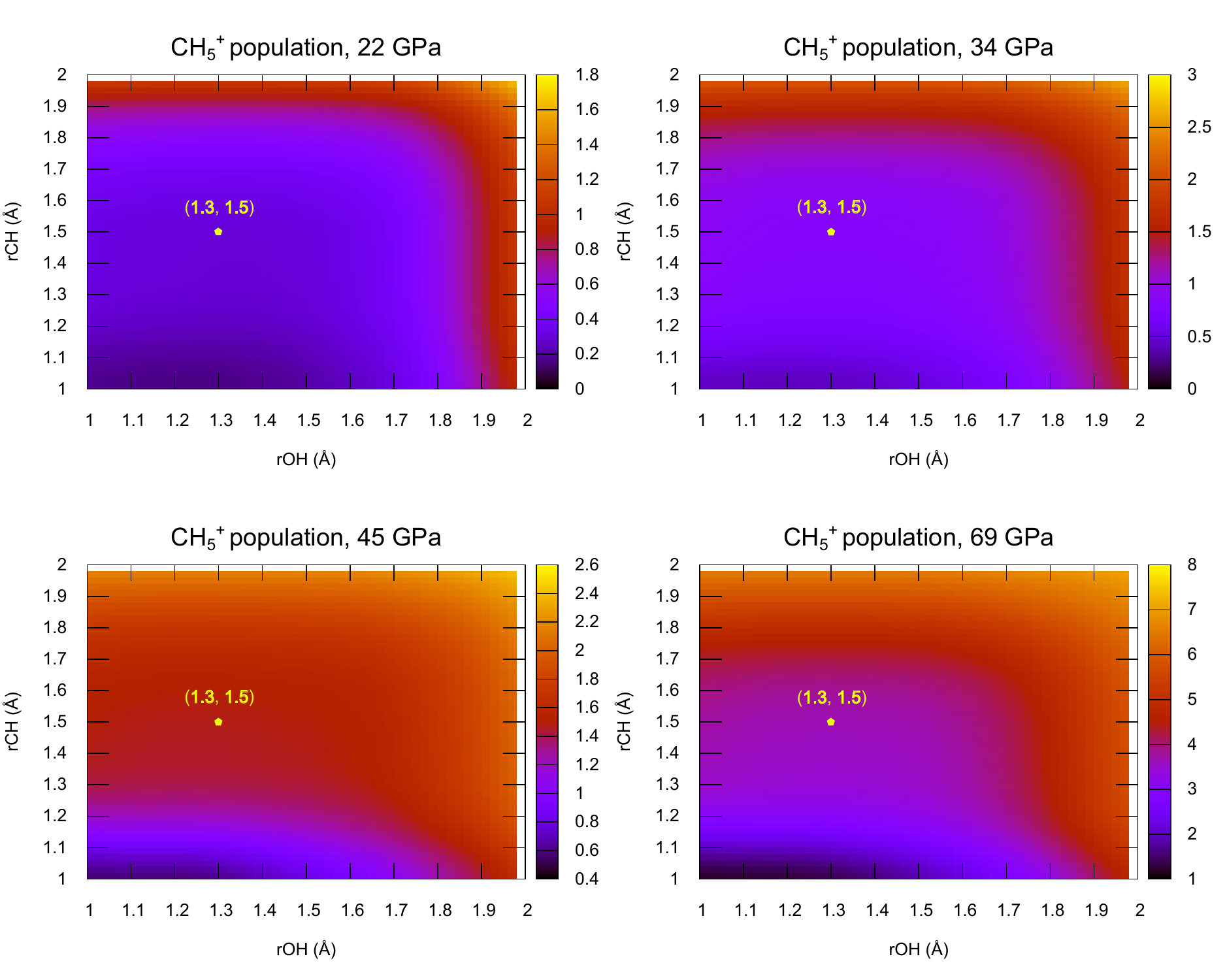}
%\hfill
\includegraphics[width=0.5\linewidth]{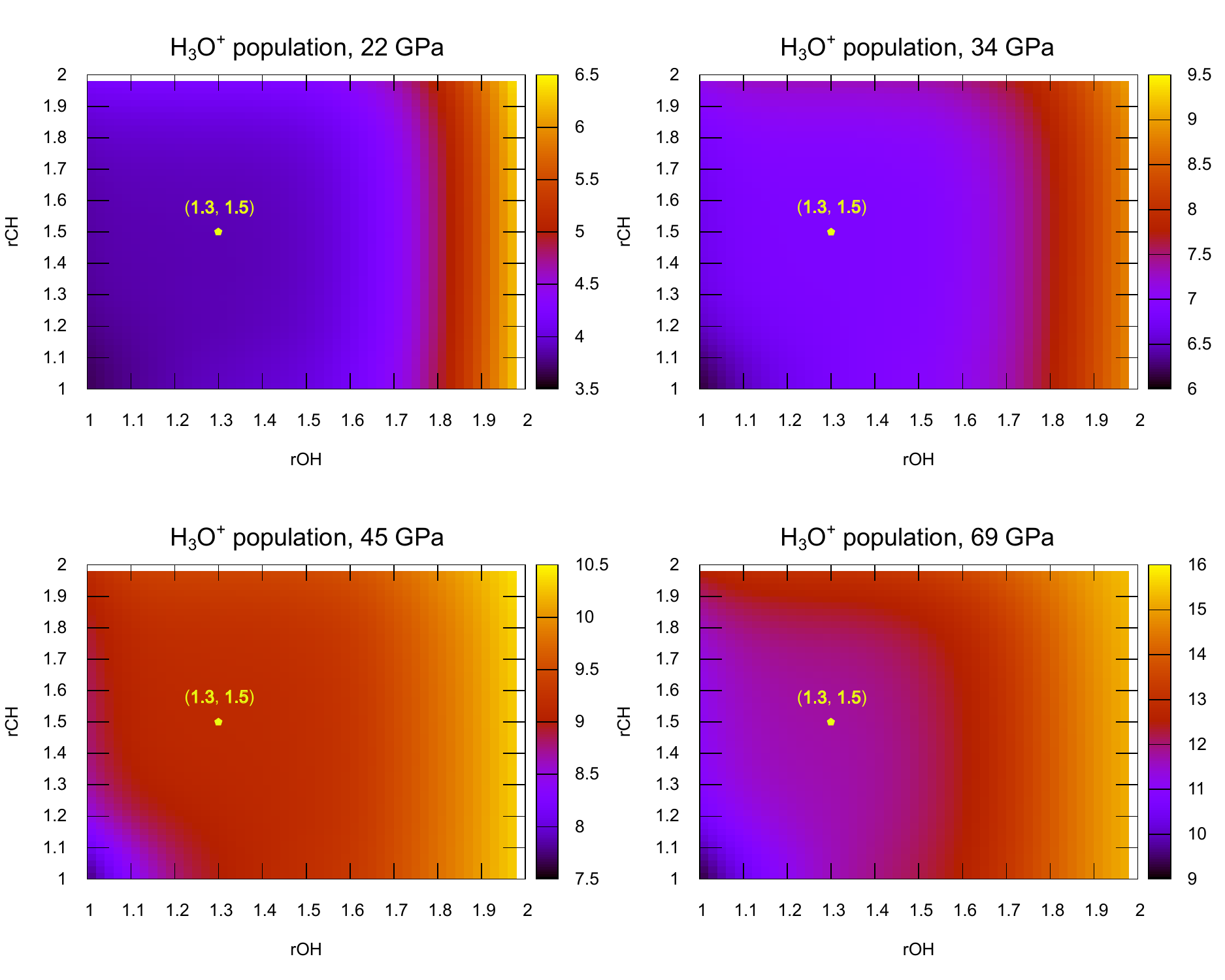}
\caption{Effect of C--H and O--H cutoff values, namely rCH and rOH, on the \ch{CH5+} (left half) and \ch{H3O+} (right half) populations in the 22-69 GPa range at 3000 K. The screening is done for cutoff values ranging from 1 to 2 \r{A}. The (rOH, rCH) point corresponding to the chosen cutoff values is reported in yellow.}
\label{fig:cutoff}
\end{figure*}

We record each instance in which a hydrogen in \ch{CH5+} experiences a nearest neighbor change from carbon to a nearby oxygen. To avoid over-counting events due to fluctuations, we exclude rapid back-and-forth proton transfers by applying an intermittent function checking for the completion of the proton transfer over a 30 fs window. Extended Data Fig. \ref{fig:tempo} shows the number of excluded proton transfers as a function of the intermittent function time. As observed, convergence is achieved from 30 fs onward.

\subsection*{DFT: labeling for MLIPs}\label{methods:dft}

We selected configurations sampled from DFT-MD and MLIP-MD to perform accurate single-point calculations. These datasets were then used to train MLIPs at different levels of accuracy - this therefore constitutes a "labeling" step. For all these calculations, we increased the SCF convergence cutoff to $10^{-7}$ Ha, and switched to TZVP-MOLOPT basis sets. We first performed calculations at the PBE level, which constitutes the "PBE large basis" dataset used later for training ("PBE small basis" being the DFT-MD reference). Then, we performed calculations at the hybrid functional level of theory, using PBE0~\cite{adamo1999toward}. We have used the FIT3 auxiliary basis set and truncated the Coulomb potential at half the system cell (6.4 \AA), following guidelines to accelerate such calculations~\cite{guidon2010auxiliary}. CP2K input scripts for both PBE and PBE0 are provided as Supplemental Material.

\subsection*{MLIPs: training}

We exclusively used the Allegro architecture~\cite{musaelian2023learning}, a strictly-local, equivariant neural network. We implemented symmetries of the SO(3) group on geometric inputs and internal features. We used a radial cutoff of 4 \AA, 64 features for irreps, 2 tensor product layers, and a maximal rotation order $l$ for spherical harmonics set to 2. Input scripts describing the selected architecture and the optimization process are included as Supplemental Material. 

We followed an iterative training strategy in which successive MLIPs are used to generate new configurations using molecular dynamics, which are then labeled using DFT. First, we sampled configurations from DFT-MD calculations at 3000 K and 45 GPa, by downsampling a 24 ps long trajectory (equilibration being excluded) using a stride of 50 fs. This amounts to 480 configurations, out of which 50 are used for validation. In a first iteration, after training a first model on the DFT dataset, we performed a single molecular dynamics simulations of 700 ps, from which we sampled 70 frames (one every 10 ps). We then labeled these configurations, added them to the original DFT dataset, and trained a new model. We repeated this operation 4 times, although we performed each time 8 independent molecular dynamics simulations of 150 ps, sampling new structures every 5 ps. In the end, this leads to a dataset of $480 + 70 + 240 \times 4 = 1510$ structures, out of which 50 are used for validation. 

Training is performed using a single MI250 GPU, with the Adam optimizer and default hyperparameters. The loss function includes both the mean squared error of the per-atom energies, and the mean squared error of the atomic forces. We adopt an initial learning rate of $10^{-5}$, and schedule its reduction by a factor of 0.35 following a plateau during 50 epochs of the loss. We select the model that minimizes the validation loss during training. On the final datasets, we obtain validation set root mean squared errors on the energies and the forces on the order of 6 meV/atom and 175 meV/\AA{} (PBE), and 7 meV/atom and 204 meV/\AA{} (PBE0). The errors on the atomic forces are relatively large compared to typical figures obtained for simpler chemical systems at atmospheric conditions. This is associated to the highly reactive environment at play; the figures obtained compare favorably to the ones reported previously for pure hydrocarbon mixtures under extreme conditions~\cite{cheng2023thermodynamics}. We have also prepared a challenging test set, composed of configurations spanning the chemical diversity observed in our molecular dynamics simulations. First, we have projected our datasets onto a $204$-dimensional space of the concentrations, in carbon and oxygen, of 102 identified chemical species. Then, using farthest point sampling, we have selected 1000 configurations corresponding to the ones showing the highest chemical diversity. Plots of learning curves and error distributions for each dataset are reported in the Supplementary Figs. 10 and 11, as well as training/validation/test error metrics in Supplementary Table 6. 

\subsection*{MLIPs: molecular dynamics}

All MLIP-based molecular dynamics simulations have been performed using \textsc{LAMMPS}~\cite{thompson2022lammps} (7 Feb 2024), patched for Allegro and for D3 dispersion corrections which we implement analytically. We use interaction and coordination number cutoffs of 11.2 \AA{} for D3 Becke-Johnson corrections. To estimate free energy profiles, we perform 8 independent 10 ps long simulations, in the $NVT$ ensemble at 3000 K, using a time step of 0.5 fs, and a Nosé-Hoover chain of three thermostats with a time constant of 50 fs. 

To take into account nuclear quantum effects, we perform normal-mode path integral molecular dynamics with the stochastic path integral Langevin equation (PILE) thermostat~\cite{ceriotti2010efficient}, using a damping time of 50 fs on the centroid mode. We use 6 polymer beads and have tested for convergence, as reported in Supplementary Fig. 12. 

\bmhead{Supplementary information}

Supplementary material is available online. 

\bmhead{Acknowledgements}

F.S.B. discloses support for the research of this work from Agence Nationale de la Recherche (grant number ANR-23-CE29-0001). A.F.L. discloses support for the research of this work from Agence Nationale de la Recherche (grant number ANR-23-CE30-0017). This work was granted access to HPC resources provided by GENCI under the allocations SS010815399, A0160814154, A0140814154. Special thanks to Dr. Daria Ruth Galimberti for fruitful discussions.

\onecolumn

\begin{table}[h!]
        \centering
        \caption{Comparison of species populations computed by molecular recognition with Wannier centers (Population 2) and without (Population 1).}
        \label{tbl:population}
        \begin{tabular}{ccc}
        \hline\\[-2ex]
            Species & Populations 1 ($\%$) & Populations 2 ($\%$) \\[.5ex]
        \hline\hline\\[-2ex]
               \ch{OH-}  & 9.518 & 9.517 \\
               \ch{H2O}  & 80.45 & 80.44 \\
               \ch{H3O+} & 9.252 & 9.261 \\
               \ch{CH4 } & 88.89 & 88.89 \\
               \ch{CH5+} & 1.491 & 1.475 \\[.5ex]
        \hline
        \end{tabular}
\end{table}

\end{appendices}

\newpage

\bibliography{library.bib}



\section*{Supplementary material}

# 1 Water dissociation degree

In Fig. 1 we report the population of $H_2O$, $H_3O^+$ and $OH^-$ species in the range 22-69 GPa at 3000K for the $CH_4/H_2O$ mixture (left) and pure liquid water (right). We observe that the presence of $CH_4$ reduces water ionization in the 22–69 GPa range at 3000 K. Interestingly, the reduction of water ionization degree with respect to its value in pure water becomes less pronounced with increasing pressure. At 22 GPa the water ionization degree exhibit a relative change of 51.9% of its value, while this effect goes down to 41.7% at 34 GPa, 30.8% at 45 GPa and 18.3% at 69 GPa.

This result is consistent with the enhancement of the methane dipole moment under rising pressure contributing to stabilizing water ionization by facilitating charge separation.

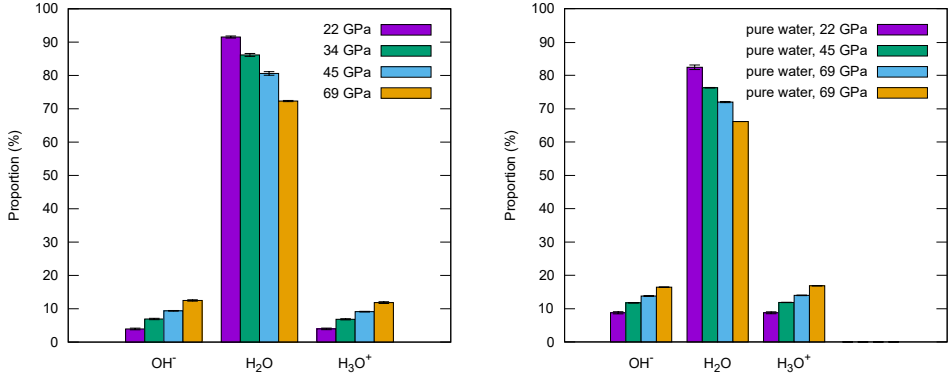

**Fig. 1** Population of $H_2O$,$H_3O^+$ and $OH^-$ species in the range 22-69 GPa at 3000K for the $CH_4/H_2O$ mixture (left) and pure liquid water (right). All values are accompanied by an error bar corresponding to a 95% confidence level. For liquid water, we used simulation boxes containing 128 water molecules (384 atoms) for each pressure, and each system was then simulated for an average of 28 ps using DFT-MD.

# 2 CH$_5^+$ lifetime distributions

In Fig. 2 we report the CH$_5^+$ lifetime distributions in the 22-69 GPa range at 3000 K. As the reader can appreciate, the lifetime distributions are not significantly affected by the increasing pressure.

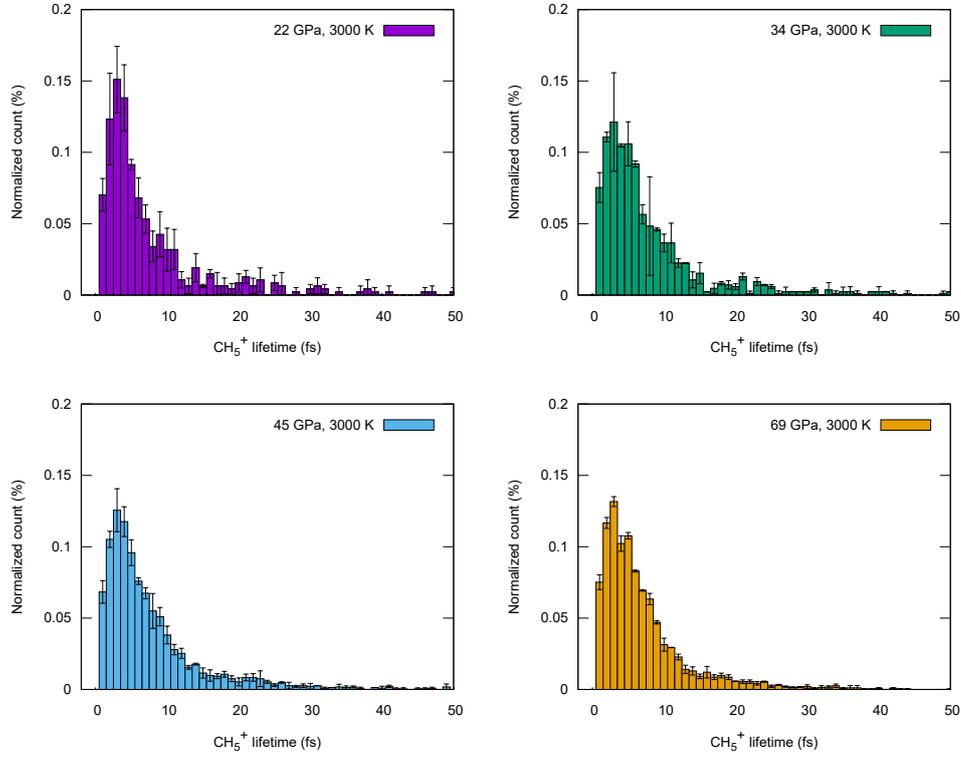

**Fig. 2** CH$_5^+$ lifetime distributions in the 22-69 GPa range at 3000 K. All values are accompanied by an error bar corresponding to a 95% confidence level.

# 3 Free energy profile for $CH_5^+$ formation from $CH_4 + H_2O$

In Fig. 3 we report the free energy profiles for the $CH_5^+$ formation from the $CH_4 + H_2O$ reaction channels at 3000 K in the pressure range 22-69 GPa. Despite the more than twofold increase in free energy barriers compared to the reaction involving methane and hydronium, the pressure-dependent decrease in free energies follows a qualitatively similar trend.

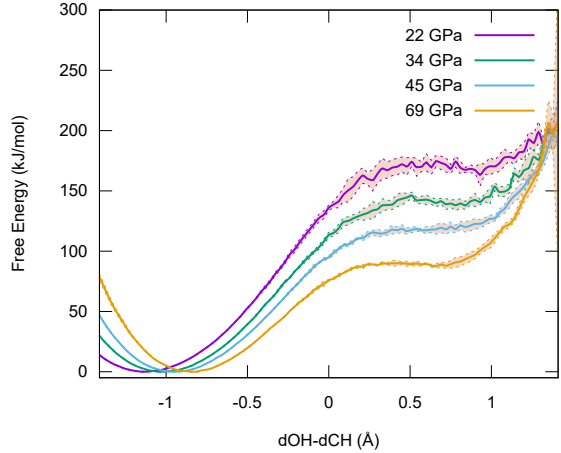

**Fig. 3** Free energy of $CH_5^+$ formation from $CH_4 + H_2O$ as a function of the reaction coordinate, defined as $d(\text{O-H}) - d(\text{C-H})$, the difference between the distance from the hydrogen to the nearest oxygen and the distance from the same hydrogen to the nearest carbon atom. All values are accompanied by an error bar corresponding to a 95% confidence level.

**Table 1** The first column displays the pressure conditions. Columns two, three and four report the diffusion coefficients at 3000 K for the different atom kinds.

| $P$ (GPa) | $D_O$ (m$^2$/s) | $D_C$ (m$^2$/s) | $D_H$ (m$^2$/s) |
|---|---|---|---|
| 22 | $2.7987 \times 10^{-8}$ | $2.1921 \times 10^{-8}$ | $2.6425 \times 10^{-8}$ |
| 34 | $1.9185 \times 10^{-8}$ | $1.3221 \times 10^{-8}$ | $2.3571 \times 10^{-8}$ |
| 45 | $1.6707 \times 10^{-8}$ | $1.4959 \times 10^{-8}$ | $2.6346 \times 10^{-8}$ |
| 69 | $0.9240 \times 10^{-8}$ | $0.7748 \times 10^{-8}$ | $2.5878 \times 10^{-8}$ |

# 4 Dipole distributions of liquid CH$_4$ *vs* CH$_4$/H$_2$O mixture

Fig. 4 shows the dipole moment distribution of pure liquid CH$_4$ (red) and the CH$_4$/H$_2$O mixture (blue). The data reveal that at 47 GPa and 3000 K, pure liquid methane has a significantly lower dipole moment than methane in the water–methane mixture under similar conditions. This confirms a significant effect of water's ionization and polarization at high pressures in enhancing methane's dipole moment.

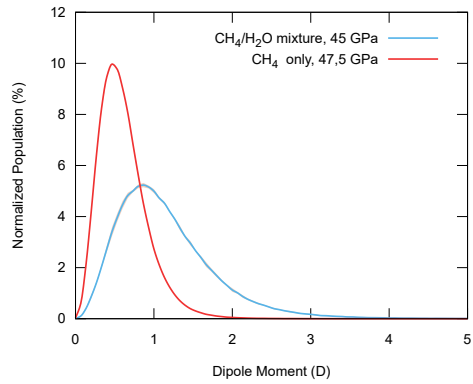

**Fig. 4** Population of the CH$_4$ molecular dipole moments of CH$_4$/H$_2$O mixtures (blue) and pure liquid methane (red) at 45 GPa and 3000K. The box of pure methane is composed by 475 atoms (95 methane molecules), simulated for 22 ps by GGA-DFT-MD. All values are accompanied by an error bar corresponding to a 95% confidence interval.

# 5 Superacid behaviour of water in 2000-3000 K range for low $CH_4$ concentrations

To investigate whether superacidity [1–7] persists in water at lower $CH_4$ concentrations and below 3000 K, we performed two independent DFT-MD simulations of a water–methane mixture composed by 1 methane and 127 water molecules at 45 GPa and at temperatures of 2000 and 3000 K, respectively. In both simulations, the methane molecule participates in proton-hopping mechanisms with water. The mechanism, along with the Wannier centers associated with the reactive structures, is illustrated in Fig. 5 for the reaction at 2000 K and 45 GPa. The methane first accepts a proton (depicted in orange) from a nearby water molecule, forming a $CH_5^+$ species (identified by its Wannier centers). Subsequently, it donates a proton (blue) to another water molecule. The structure labeled as II clearly exhibits the two-electron three-center bond [8–15] characteristic of the $CH_5^+$ species. This process occurs after 0.5 ps at 3000 K and after 2.5 ps at 2000 K. Our results demonstrate that water maintains its superacidic behavior even at 2000 K and 45 GPa, despite the relatively low concentration of $CH_4$.

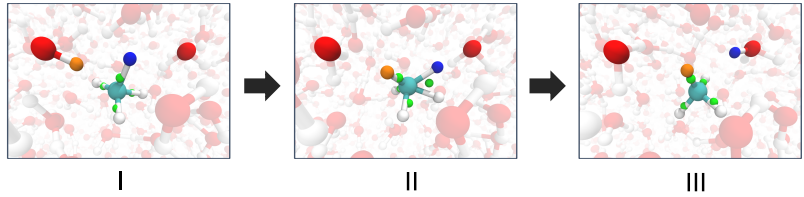

**Fig. 5** Reaction structures with Wannier centers associated to the Grotthuss-like proton-hopping mechanism between water and methane at 45 GPa and 2000 K. The proton donated by water is shown in orange, while the proton donated by methane is shown in blue.

# 6 M3 mechanism and lifetime distributions of ethene and formaldehyde

The M3 mechanism is an elongation reaction involving the formation of double-bond intermediates. In this process, the double-bonded species acts as an electrophilic partner to a neighboring molecule. The M3 mechanism can proceed via either a base-catalyzed or a superacid-catalyzed pathway. In the superacid M3 mechanism (presented in Fig. 3c of the main text), the double bond intermediate is produced by a hydrocarbon or alcohol cation species formed through a superacid mechanism. By contrast, in the base-catalyzed mechanism, the double bond intermediate derives from an anionic species, previously generated by deprotonation of hydrocarbon or alcohol from a base. In Fig.6, we illustrate the spontaneous formation of ethanol via the M3 base-catalyzed pathway, as observed in an unbiased MD simulation at 3000 K and 45 GPa. The process begins with the deprotonation of methanol at the O–H bond by a hydride ($H^-$) from methane, that acts as the base. The resulting $CH_3O^-$ evolves, forming a C=O double bond, that leads to the formation of formaldehyde and the release of hydride ($H^-$) that recombines with $CH_3^+$. Subsequently, a methane molecule in the proximity performs an electrophilic attack on the formaldehyde. This results in the breaking of the C=O bond and the elongation of the alcohol chain, leading to an ethanol molecule (Structures VI, Fig.6). It is important to note that the M3 base mechanism was also characterized in metadynamics DFT-MD simulations, where also $H_2O$ or $OH^-$ have been found to play the role of base in the initial deprotonation stage.

In Fig. 7, we present the lifetime distributions of ethene (top) and formaldehyde (bottom), the two M3 intermediates observed in both superacid and base catalyzed variants, at 3000K under varying pressure conditions. The distributions clearly demonstrate a decrease in intermediate stability with increasing pressure, as evidenced by the shift toward shorter lifetimes. This behavior highlights the reduced stability of $\pi$ bonds at elevated pressures, which can be attributed to their relatively large spatial occupancy.

## M3 mechanism

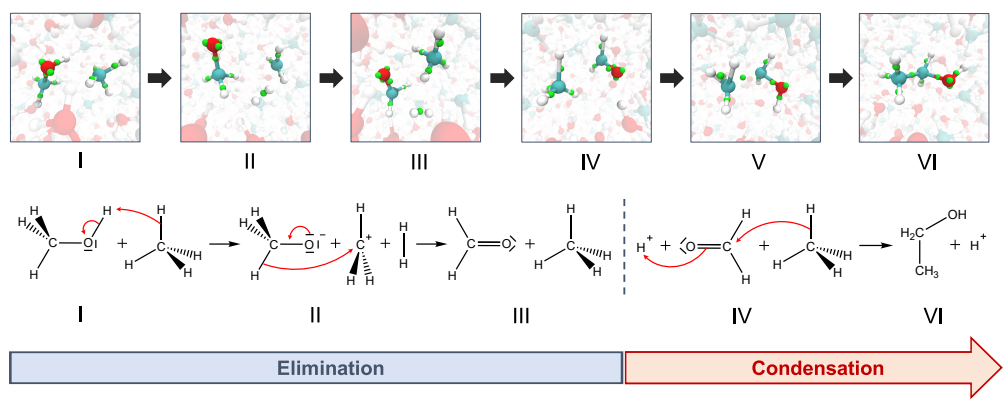

**Fig. 6** Scheme of the reaction mechanisms, along with the Wannier centers associated with the reactive structures for the M3 base-catalyzed mechanism.

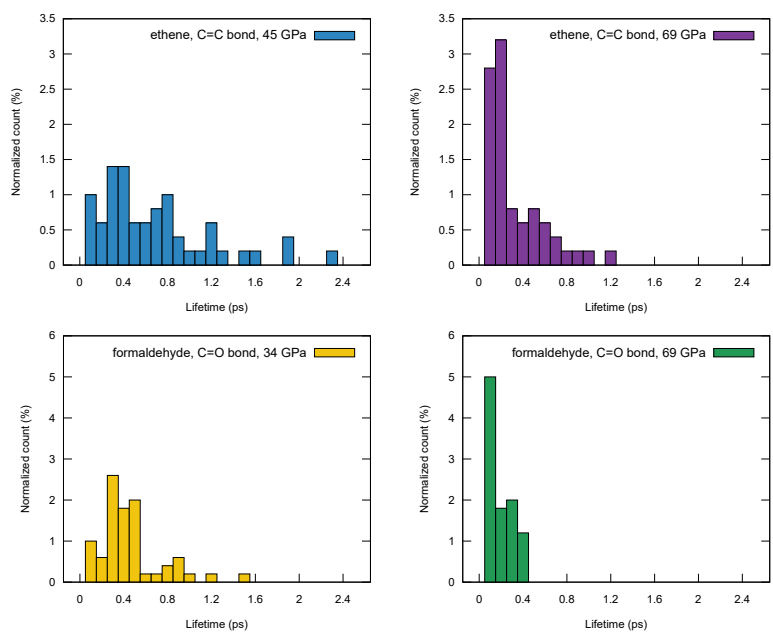

**Fig. 7 Top Panel**: Lifetime distributions of ethene ($C_2H_4$) at 45 GPa (blue, left) and 69 GPa (purple, right) at 3000 K. **Bottom Panel**: Lifetime distributions of formaldehyde ($CH_2O$) at 34 GPa (yellow, left) and 69 GPa (green, right) at 3000 K. Each distribution is based on 50 DFT-MD simulations using a 488-atom simulation box, totaling 200 DFT-MD simulations across all pressures.

# 7 Formation of diamond-like structure along unbiased DFT-MD

In Fig. 8, we follow the transformation from a ternary to a quaternary hydrocarbons species, by computing the coordination numbers of the reactive atoms along the unbiased DFT-MD simulation at 45 GPa and 3000 K. In particular, the coordination numbers of the hydrocarbon central carbon with respect to all the carbons (black curve) and oxygen atoms of the system (red curve) are reported as a function of time.

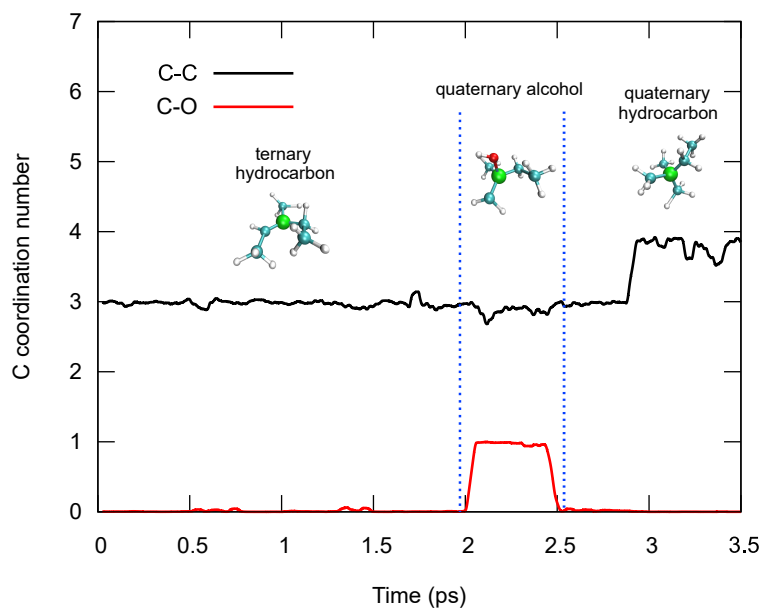

**Fig. 8** Evolution of the coordination number of the hydrocarbon central carbon (in green) during its spontaneous transformation from a ternary to a quaternary species in an unbiased MD simulation. The red line depicts the C–O coordination number, while the black line shows the C–C coordination number.

# 8 SPRINT metadynamics

The SPRINT coordinates [16, 17] are defined as:

$$S_i = \sqrt{N} \times \lambda^{\max} \times v_i^{\max,\text{sorted}}, \quad i = 1, 2, ..., N \tag{1}$$

where $N$ is the number of atoms, $\lambda^{\max}$ and $v^{\max,\text{sorted}}$ are the largest eigenvalue and corresponding eigenvector (with sorted entries) of the smooth adjacency matrix of atoms, composed by $a_{ij}$ elements:

$$a_{ij} = \frac{1 - (r_{ij}/r_0)^n}{1 - (r_{ij}/r_0)^m} \tag{2}$$

where $r_{ij}$ are interatomic distances and $r_0$, n, and m are parameters depending on the typical bond lengths of the system under consideration. The parameters for the set of switching functions, specifically the $r_0$, $n$, and $m$ parameters, are provided for all atom pairs in Table 2.

**Table 2**

| Atoms | $R_0$ (Å) | N | M |
|-------|-----------|---|---|
| C-C | 2.40 | 4 | 16 |
| C-H | 1.10 | 8 | 16 |
| C-O | 2.70 | 8 | 12 |
| H-H | 0.74 | 8 | 16 |
| H-O | 0.95 | 8 | 16 |
| O-O | 1.40 | 8 | 16 |

The $S_i$ SPRINT variables, obtained by diagonalizing the smooth interatomic adjacency matrix, are centered on each atom of the system and capture changes in both its short- and long-range connectivity. We have benefited from the dimensional reduction allowed by the magnitude-ordering, which introduces correlations among the values of the $S_i$. For this reason, in between 16 and 24 out of the SPRINT variables were biased in our metadynamics simulations. Table 3 provides details of the SPRINT simulations performed on a $CH_4/H_2O$ box containing 732 atoms, which led to the formation of a diamond-like structure (see reaction mechanism in Fig. 4b of the main text). The table lists the simulation lengths, the maximum value of the metadynamics bias, the metadynamics parameters, the number of biased SPRINT variables ($N_{SPRINT}$) and the heaviest hydrocarbon species formed during the metadynamics. Similarly, Table 4 summarizes the same parameters for six independent SPRINT simulations from box containing 488 atoms, which resulted in the formation of complex hydrocarbon species (see reaction mechanism for 3-methyl pentane formation in Fig. 9). For all the metadynamics the Hills deposition time is of 100 steps. The *ab initio* MD simulations, have been coupled with the metadynamics algorithm [18–21] via the Plumed plugin [22].

**Table 3** Parameters for the SPRINT metadynamics simulations leading to the formation of a quaternary diamond-like structure from a simulation box containing 722 atoms, consisting of one 3-methylpentane molecule in a $CH_4/H_2O$ mixture at 3000 K and 45 GPa. The first column reports the simulation time, while columns two through six provide details on the metadynamics hill height, hill width, the number of biased SPRINT coordinates, and the heaviest hydrocarbon species formed.

| Time (ps) | Maxium bias ($k_BT$) | Hills height ($k_BT$) | Hills width | $N$ | Hydrocarbon |
|---|---|---|---|---|---|
| 7.8 | 16 | 1.4 | 2.5 | 24 | 3,3-dimethylhexane |

**Table 4** Parameters for the SPRINT metadynamics simulations leading hydrocarbon elongation and branching from $CH_4/H_2O$ systems containing 488 atoms at 3000 K and 45 GPa. The first column lists the simulation time. Columns two through six detail the metadynamics hill height, hill width, the number of biased SPRINT coordinates, and the heaviest hydrocarbon species formed.

| Time (ps) | Maxium bias ($k_BT$) | Hills height ($k_BT$) | Hills width | $N$ | Hydrocarbon |
|---|---|---|---|---|---|
| 9 | 27.7 | 1.4 | 2.5 | 16 | 3-methylpentane |
| 17 | 27.8 | 1.4 | 1.5 | 22 | 2-methylpentane |
| 17 | 21.9 | 1.4 | 2.5 | 24 | pentane |
| 16 | 17.2 | 0.4 | 2.5 | 24 | propane |
| 15 | 29.0 | 1 | 2.5 | 24 | propane |
| 14 | 23.3 | 1.4 | 2.5 | 24 | propane |

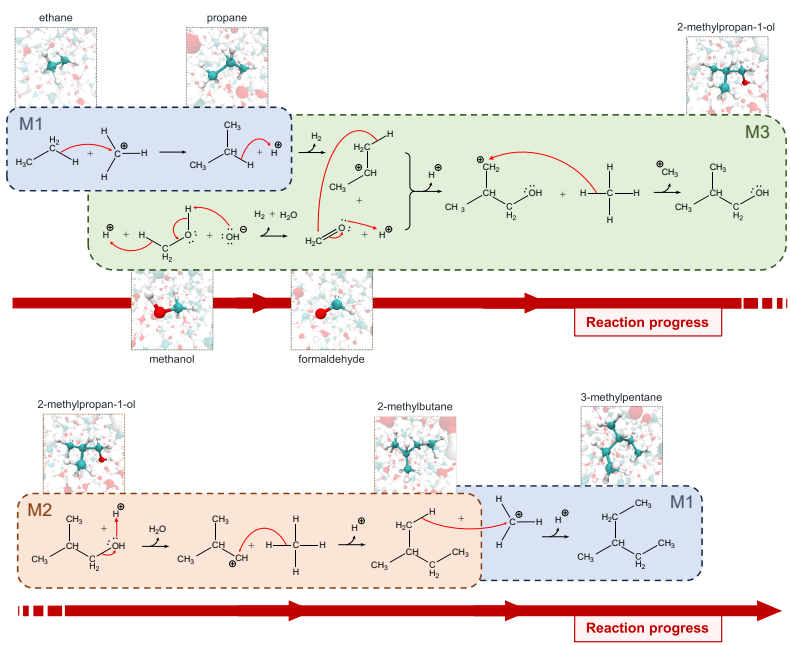

**Fig. 9** Scheme of the reaction mechanism for 3-methylpentane formation, containing a ternary hydrocarbon structure, at 3000K and 45 GPa obtained by SPRINT-based hydrocarbons simulation. The superacid M1, M2 and a base-catalyzed M3 mechanisms, previously characterized by unbiased DFT-MD simulations, are identified.

# 9 Machine learning interatomic potentials

Many exotic phenomena observed *in silico* are artifacts due to poorly converged simulations, or inappropriate approximations [23]; it is therefore mandatory to validate our observations on superacid chemistry. Since GGA functionals are known for their inaccuracies on reaction barriers and energies [24], we have evaluated the consistency of our results using a hybrid functional. In addition, we tested the possible influence of basis set size, as well as nuclear quantum effects. For such tests to be affordable, we have implemented surrogate models at 45 GPa in the form of machine-learned interatomic potentials [25] (MLIPs), their training starting from the configurations sampled through DFT-MD. The results are presented in the main text, in Fig. 2e,f. First, we demonstrate that our surrogate model can reproduce the thermodynamics and kinetics of proton transfer involving $H_3O^+$ and $CH_5^+$: the $\Delta F$ and $\Delta F_1^\ddagger$ predicted by the surrogate model and DFT only differ respectively by less than 1 and 0.3 $k_BT$. Second, nuclear quantum effects show only negligible effects, with reactions favored by less than 0.1 $k_BT$. Third, basis set size is shown to have no effect on the reaction we monitor, with both free energy profiles associated with small and large basis sets being superimposed. Finally, we switch to a hybrid functional (PBE0). We must assume that training against hybrid or GGA datasets leads to a comparable error on the free energy profile, since we do not have access to a direct hybrid reference. Given the similarities between the error metrics and distributions (see Supplementary Fig. 10 and Tables 5 and 6), this seems to be a reasonable assumption. Switching to a hybrid functional has a slightly larger effect, since the $CH_5^+$ state is destabilized by less than $0.5k_BT$, and the barrier increases by $0.6k_BT$. The effects remain however marginal, since the absolute barrier at the hybrid level is low (less than $2k_BT$).

We report here the training, validation, and test metrics obtained for all three models used (Tables 5 and 6). In addition, we present error distributions on the atomic forces as a function of the norm of the target forces (Fig. 10), an example of optimization result for the "PBE - small basis" model (Fig. 11), and the convergence of free energy profiles with respect to the number of beads used in our normal-mode path integral molecular dynamics simulations (Fig. 12).

| MAE / RMSE (meV/Å) | Train | Validation | Test |
|---|---|---|---|
| PBE - small basis | 122 / 171 | 122 / 181 | 142 / 215 |
| PBE - large basis | 114 / 159 | 118 / 175 | 136 / 206 |
| PBE0 | 125 / 175 | 132 / 204 | 156 / 246 |

**Table 5** Force errors on training, validation and test datasets.

| MAE / RMSE (meV/atom) | Train | Validation | Test |
|---|---|---|---|
| PBE - small basis | 11.6 / 14.9 | 5.0 / 6.3 | 4.9 / 6.0 |
| PBE - large basis | 12.3 / 15.5 | 4.6 / 6.1 | 7.4 / 8.8 |
| PBE0 | 12.4 / 15.3 | 5.9 / 7.4 | 7.9 / 9.8 |

**Table 6** Per-atom energy errors on training, validation and test datasets.

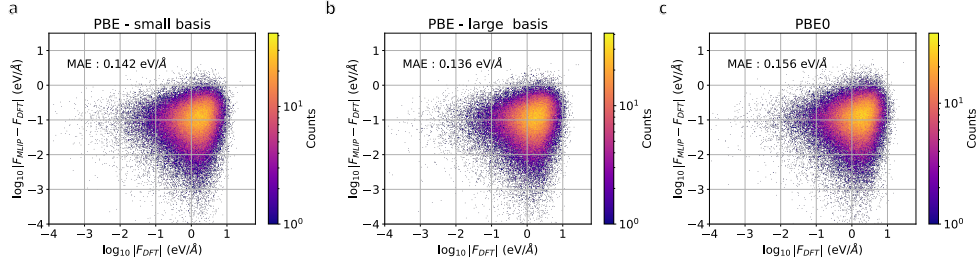

**Fig. 10** Test set atomic force error distributions as a function of the reference force norms for the three MLIPs implemented (**a**: "PBE - small basis", **b**: "PBE - large basis", **c**: "PBE0") in logarithmic scale.

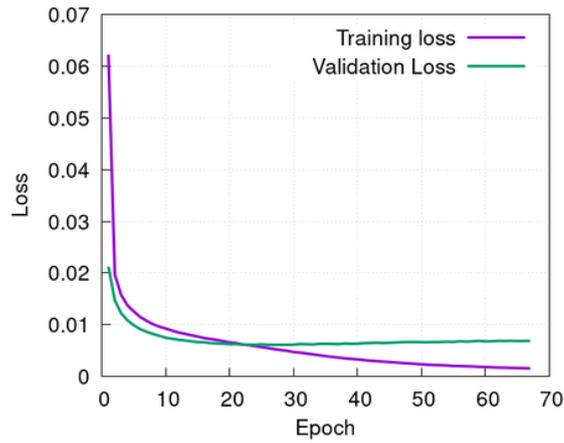

**Fig. 11** Evolution of the training and validation loss during optimization for the "PBE - small basis" dataset.

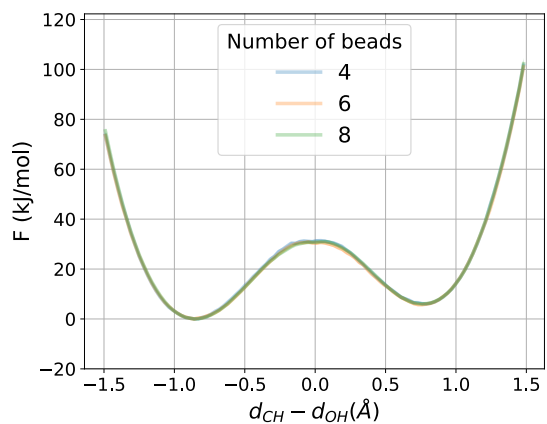

**Fig. 12** Free energy profile associated with the proton transfer from $H_3O^+$ to $CH_5^+$ obtained from normal-mode path integral molecular dynamics simulations, with either 4, 6, or 8 polymer beads.



\end{document}